\documentclass[aps,pra,twocolumn,amsfonts,amssymb,epsfig,superscriptaddress,showpacs]{revtex4}

\usepackage{graphics,graphicx}
\usepackage{amsmath,bbm}
\usepackage{amstext}
\usepackage{amssymb}

\def\beq{\begin{equation}}
\def\eeq{\end{equation}}
\def\bea{\begin{eqnarray}}
\def\eea{\end{eqnarray}}
\def\Tr{\mathrm{Tr}}
\providecommand{\openone}{\leavevmode\hbox{\small1\kern-3.8pt\normalsize1}}

\begin{document}

\title{Critical exponents of one-dimensional quantum critical models by means of MERA 
tensor network}

\author{S. Montangero}
\affiliation {NEST CNR-INFM \& Scuola Normale Superiore, Piazza
dei Cavalieri 7, I-56126 Pisa, Italy}
\author{M. Rizzi}
\affiliation{Max-Planck-Institut f\"{u}r Quantenoptik, Hans-Kopfermann-Str. 1, D-85748 Garching, Germany}
\author{V. Giovannetti}
\affiliation {NEST CNR-INFM \& Scuola Normale Superiore, Piazza
dei Cavalieri 7, I-56126 Pisa, Italy}
\author{Rosario Fazio}
\affiliation {NEST CNR-INFM \& Scuola Normale Superiore, Piazza
dei Cavalieri 7, I-56126 Pisa, Italy}
\affiliation {International School for Advanced Studies (SISSA),
 Via Beirut 2-4, I-34014 Trieste, Italy}

\date{\today}

\begin{abstract}
An algorithm for optimizing the MERA tensor network 
in an infinite system is presented. Using this technique  we
compute the critical exponents of Ising and XXZ model.
\end{abstract}

\pacs{03.67.-a,05.30.-d,89.70.-a}

\maketitle

\section{Introduction}
Critical phenomena are ubiquitous in science ranging from condensed matter to biological 
or economic systems. They are associated to scale invariance, a diverging correlation lenght, 
and the various correlation functions decay as power law. A key issue in the study of 
of critical systems is the computation of the critical exponents, i.e. the exponents which 
govern the decay of the correlations.  The power of methods like the renormalization group 
stems from its capability to compute (analitically or numerically) the critical exponents. 

In the field of quantum many-body systems important progresses have been made recently in 
one-dimension where a number of tensor network approximations~\cite{FNW,OR,VIDAL1,VPC,rev,
peps,wgs,mera} of the many-body wave-function allowing an efficient computation of  all 
the  relevant observables. The Multi-scale Entanglement Renormalization Ansatz (MERA)
introduced by Vidal~\cite{mera} is particularly appealing since embeds  
scale invariance typical of critical systems. A method to relate the critical exponents 
in the MERA tensor network has been very recently put forward by some of us~\cite{quMERA}
by relating them to the eigenvalues of the MERA transfer matrix.

In the present work we want to apply this method for the computation of the critical 
exponents of two cornerstone models in quantum statistical mechanics, the Ising and the 
XXZ model. To this end we provide an algorithm for optimizing the MERA tensor network 
in an infinite system and then, by employing the results given in~\cite{quMERA} we 
compute the critical exponents.

The paper is organized as follows. In the next section we briefly review the optimization of 
the MERA tensor network and how to compute the critical exponent. We then proceed with the 
comparison with the Ising and XXZ model defined in Section~\ref{model}. In Section~\ref{results}
we discuss the results and compare them with the exact known values. The last Section is
devoted to the conclusions.

\section{MERA optimization}
\label{minimization}
 
 Consider a $1$-D translational invariant many-body quantum system at criticality 
composed of $L=2^{\ell+1}$ sites, 
each of them described by a local Hamiltonian and some nearest
neighbor interactions. 
The global Hamiltonian can  then be expressed as
${H}= \sum_{k} {h}_{k}$,
where $k=1,\cdots, L$ with $L+1 \equiv 1$ for periodic
boundary conditions and where $h_k$ is acting on the three consecutive sites
$k-1$, $k$, and $k+1$
(the generalization to longer range interaction is straightforward).
In the {\em thermodynamic limit}  of infinitely many sites 
the ground state energy per site of the system can thus be computed as
\bea
E_G:=\lim_{L\rightarrow \infty} \tfrac{ \min_{\Psi} 
\langle \Psi | H |\Psi \rangle}{L}  \label{genergy}=\lim_{L\rightarrow \infty} \tfrac{
 \min_{{\cal B}_\Psi}
\sum_{k=1}^L 
 \Tr[{\rho}_k\;  h_{k}]}{L},
\eea 
where $|\Psi\rangle$ are joint states of the many-body system, and where 
the last minimization is performed over the sets 
${\cal B}_\Psi:=\{ \rho_1, \cdots, \rho_L\}$ whose elements  can be obtained 
as reduced density matrices  of some global pure state $|\Psi\rangle$ of the system
(specifically $\rho_k$ is the reduced density matrix of $|\Psi\rangle$  that is associated with
the sites $k-1$, $k$ and $k+1$).
Owing to translational invariance all the addends of the last term of 
Eq.~(\ref{genergy}) are equal. Thus without the constraint $\rho_k\in{\cal B}_\Psi$ 
the minimization would be trivial and $E_G$ will coincide with the 
the ground energy level of the 3-sites Hamiltonian $h_k$.
Taking into account the condition $\rho_k\in {\cal B}_\Psi$ 
is what makes the calculation of $E_G$ a
{\em hard} problem to solve: it requires us to minimize the energy of  $h_k$ by  
 properly embedding $\rho_k$ in the "environment" formed by the remaining sites of the many-body system. 
 
A solution  can be found by  adopting the so called 
Multi-scale Entanglement Renormalization 
Ansatz (MERA)~\cite{mera}, which,
 similarly to the matrix product state decomposition~\cite{FNW,OR,VIDAL1,VPC,rev}, 
  assumes a specific representation of the many-body state $|\Psi\rangle$. 
 In particular according to this ansatz
the wave function 
of  an $L$ sites many-body system
 is expressed in terms of $O(L \log_2 L)$ tensors 
 of two different species 
 (i.e. the type-$
 \mbox{\tiny{$\left(\begin{array}{c} 2 \\2
 \end{array}\right)$}}$ {\em disentangler} tensors $\chi$
and type-$
 \mbox{\tiny{$\left(\begin{array}{c} 1 \\2
 \end{array}\right)$}}$ {\em isometry} 
 tensors ${\lambda}$) which are connected to form a multi-layer 
 structure 
which admits efficient contraction rules 
(see Fig.~\ref{oned} --- we refer the reader to Ref.~\cite{mera} for details).
Specifically, following Ref.~\cite{quMERA}
 we will restrict the minimization~(\ref{genergy}) to 
states $|\Psi\rangle$ which can be expressed (or at least approximated) by {\em homogenous} MERAs,  in which all  the  disentangler $\chi$  and all the isometry  $\lambda$ entering the tensor network
are identical.
Under these conditions it has been shown~\cite{quMERA} that 
the thermodynamic limit $L\rightarrow \infty$ of $|\Psi\rangle$ can be
 characterized in terms of a transfer super-operator (the QuMERA map $\Phi$) that is  determined
by the $\chi$s and the $\lambda$s that forms the MERA~\cite{NOTA11}.
In particular 
the reduced density matrix $\rho_k$ of $|\Psi\rangle$ can  be computed as the 
(unique) eigenvector $\rho_k^*$ of the MERA transfer super-operator (QuMERA map) $\Phi$ associated with the unitary eigenvalue,  i.e.
\beq
\Phi(\rho_k^*) = \rho_k^*\;.
\label{eigen}
\eeq
This means that Eq.~(\ref{genergy}) can now be expressed as
\bea
E_G= \mathrm{min}_{{\Phi}} \; \Tr[ \rho_k^* \; h_{k}]\;,
\label{problem}
\eea
where the minimization  is performed over the set of all possible QuMERA channels $\Phi$
-- the equivalence being guaranteed by the stationary condition~(\ref{eigen}).
Equation~(\ref{problem}) achieves two fundamental goals:
{\em i)} it allows us to directly address the thermodynamic limit, 
{\em ii)} it guarantee the possibility of reconstructing a many-body joint state $|\Psi\rangle$
associated with the three sites local density operator $\rho_k^*$ (this is the MERA state 
corresponding to the QuMERA channel $\Phi$).
In other words, exploiting the above derivation 
the constrained energy  minimization problem~(\ref{genergy}) corresponds to  
minimize the functional
${F}(\Phi, \rho) =  \Tr[ \Phi (\rho)\;  h] + \mathcal{ L } \; 
\| \Phi (\rho) - \rho \|$,
where we omitted the index $k$ as, from now on, all the quantites refers to a
subsystem composed by three sites and $\mathcal{ L }$ 
is a Lagrange multiplier. 
This is the main result of this paper and as we will
show in the following, this allows us to compute the thermodynamic
ground state energy employing a number of  
 steps which {\em do not depend} upon the system size 
(the limit $L \to \infty$ being already included in the QuMERA description).
Once the transfer operator has been found via minimization, as shown
in~\cite{quMERA}, the long range properties (critical exponents) can
be easily computed.

\subsection{Minimization}

\begin{figure}[t!]
\begin{center}
\includegraphics[scale=0.17]{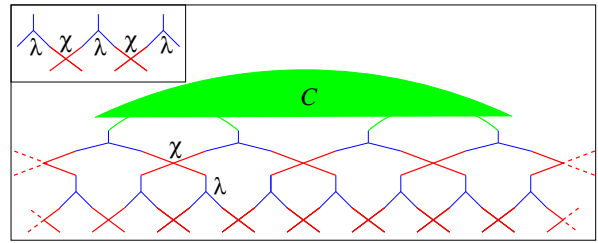}
\caption{Graphical representation of a one dimensional MERA tensor network for a  many-body system of $L=8$ sites. The red elements 
correspond to the disentaglers tensors $\chi$, while the blue elements are the isometry tensors $\lambda$
(the green element ${\cal C}$ is the {\em hat} of the MERA which plays no role in the $L\rightarrow \infty$ limit).
Any two joined legs from any two distinct nodes indicate saturation of the associated indices of the corresponding
tensor~\cite{mera}. In the inset it is shown the tensor compound which determines the transfer super-operator $\Phi$ 
associated with the MERA (see Ref.~\cite{quMERA} for details).   
} \label{oned}
\end{center}
\end{figure}

The problem of finding the ground state properties have been
reformulated in terms of the minimization of the quantity~(\ref{problem})
with the additional constrain~(\ref{eigen}).
To solve it we proceed with an iteration procedure
starting from an initial guess  
$\Phi_0$ and $\rho^*_0$ for the QuMERA channel $\Phi$ and its associated eigenvector $\rho^*$. 
Specifically, all the $\chi$ and $\lambda$ tensors 
entering in the definition of the transfer super-operator $\Phi$ are given by an initial guess $\chi_0$ and $\lambda_0$ which we  optimize by 
varying each of them  one at the time. 
In this framework the problem is  reduced to perform the following  minimizations 
\beq\label{fd1}
\min_{\chi} \left\{ \Tr \left[ 
(\lambda_0 \chi \lambda_0 \chi \lambda_0)^\dag \; \rho_0^* 
(\lambda_0 \chi \lambda_0 \chi \lambda_0) \; h
\right] 
\right\}\;,
\eeq
where the tensors are contracted following rule given in Fig.~\ref{oned}
(here we specialize in the case of the optimization of $\chi$, while
keeping fix  the initial guess for $\lambda_0$ and $\rho^*_0$). 
Possible strategies to solve such minimization have been presented
in~\cite{vidallong} where {\it unconstrained quadratic
optimization} 
have been considered.
This approach is not longer applicable in our case where the homogeneity constraint 
adopted to deal with  the translational invariance of the system, forces
 all the $\chi$s and $\lambda$s entering in Eq.~(\ref{fd1}) to be identical while  transforming the
minimization in a non-quadratic optimization problem. 
 To cope with this 
we approach Eq.~(\ref{fd1})
by computing 
at each step the linearized gradient on a basis 
for the tensors belonging to the MERA, while
alternating moving from the $\chi$s to the $\lambda$s. Once the
gradient is computed we move the tensor in a random direction whose
signs in a given basis (going forward or backward along a given
direction) are defined by the gradient's ones as in Ref.~\cite{sandvik}.

Once the optimization of the $\chi$ and the $\lambda$ have been performed and a new QuMERA channel 
$\Phi_1$ is defined, the stability constraint~(\ref{eigen}) have to be
fullfilled. Hence, we replace $\rho_0^*$ with a new guess $\rho^*_1$ 
defined as the solution of the 
eigenproblem
$\Phi_1(\rho_1^*) = \rho_1^*$,
which can be  addressed 
  by using some smart eigen-problem solvers as 
the Lancsoz or Davidson algorithms. The latter approach is favorable also due to
the presence of a ``good guess'' $\rho_0^*$ for the new eigenvector
$\rho_1^*$. The result presented in the next sections are obtained
using this second method.  
We can then proceed again in optimizing another tensor belonging to 
the transfer super-operator up to the desired convergence.
Notice that this procedure is not garanteed to converge to the optimal
minimun, however in general gives good results.

\section{The models}
\label{model}
The models we consider are defined through  the Hamiltonian 
\begin{eqnarray} 
     \label{eq:spinbath}
     {H} & = & -\frac{J}{2} \sum_j \left[
     \left( 1 + \gamma \right) \sigma^x_j \sigma^x_{j+1} +
     \left( 1 - \gamma \right) \sigma^y_j \sigma^y_{j+1} \right. \nonumber \\
     & & + \left. \Delta \sigma^z_j \sigma^z_{j+1} +
     2 \; {b} \;  
      \sigma^z_j \right] \, ,
\end{eqnarray}
where $\sigma^\alpha_i$ ($\alpha = x,y,z$) are the Pauli matrices
of the $i$-th spin. The constants $J$, $\Delta$, $\gamma$ and ${b}$ respectively characterize
the interaction strength between neighboring spins, the anisotropy
parameter along $z$ and in the $xy$ plane,
and an external transverse magnetic field.
The Hamiltonian~\eqref{eq:spinbath} has
a very rich structure~\cite{sachdev99}.  We consider 
the two cases cases:
\begin{itemize}
\item the Ising-model in a transverse field.
Here one has $\Delta =0$ and  $\gamma=1$. The model presents
a critical point at
$|{b}| = 1$.
\item the $XXZ$ anisotropic Heisenberg model.
Here one has ${b}, \gamma=0$ and $\Delta$ generic. 
In this case the Hamiltonian~\eqref{eq:spinbath} is critical for
$-1 \leq \Delta \leq 1$ while it has ferromagnetic or 
anti-ferromagnetic order for $\Delta>1$ or $\Delta<-1$ respectively. 
\end{itemize}
The exact critical exponents, related to the correlation functions 
\begin{equation}
\langle \sigma_\alpha^i \sigma_\alpha^{i+\ell} \rangle - \langle
\sigma_\alpha^{i} \rangle \langle
\sigma_\alpha^{i+\ell} \rangle 
   \sim  r^{-\nu_{\alpha}} \; ,
\end{equation}
are given by 
\begin{itemize}
\item Ising model ($\gamma=1, \Delta=0$)
\begin{equation}
\nu_z=2\;, \quad \nu_x=0.25\;, \quad \nu_y=2.25\;, 
\end{equation}
\item $XXZ$ model ($\gamma=0$, $|\Delta| \le 1$)
\begin{equation}
 \nu_x=\nu_y=1/\nu_z=1 - \mathrm{arcos}(\Delta)/\pi \;.
\end{equation}
\end{itemize}
We are going to compare our numerical results against these values.

\section{Results}
\label{results}

The algorithm we developed for solving the minimization of Eq.~(\ref{problem})
requires  to store order of $O(m^6)$ tensors with $m$ being the 
 dimension of the  indexes of the $\chi$s and $\lambda$s,  
their moltiplication ($O(m^{10})$ operations) and their diagonalization.  
As for the t-MERA~\cite{tMERA} this is ``reasonable'' up to $m=4,6$.

\begin{figure}[t!]
\begin{center}
\includegraphics[scale=0.35]{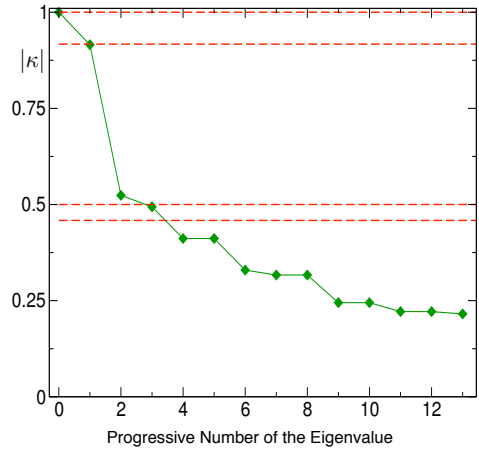}
\caption{Modulus of the eigenvalues of $\Phi$  for the Ising model for ${b}=1$, $J =1$.
The dashed lines correspond to the theoretical values obtained via Eq.~(\ref{theory}). }
\label{Ising-bmark}
\end{center}
\end{figure}

We first concentrate on the Ising model, where we have obtained an
energy convergence to the exact energy per site  of the ground state at the
thermodynamic  limit of the order of $\delta E =10^{-4}$.
In Fig.~\ref{Ising-bmark}  we report  (in decreasing order) the modulus of eigenvalues of
the QuMERA map $\Phi$ found by means of
the minimization strategy we rpesented in the last section. 
As discussed in~\cite{quMERA}  we expect the first eigenvalue ($\kappa=1$) to be 
non-degenerate as the 
$\Phi$ should be mixing.  The subsequent eigenvalues instead
 express the critical
exponents of the system via the relation:
\beq
|\kappa_\alpha^{\mathit th}|= 2^{- \nu_\alpha/2}.
\label{theory}
\eeq
The dashed red lines of the Figure report the theoretical expectations of the
eigenvalues. The result is
 good, the first one is exact up to the machine
precision, while the others have numerical errors of the orderd of 
$\delta |\kappa_\alpha| \sim 10^{-4}, 10^{-2}, 10^{-2}$.

\begin{figure}[t!]
\begin{center}
\includegraphics[scale=0.4]{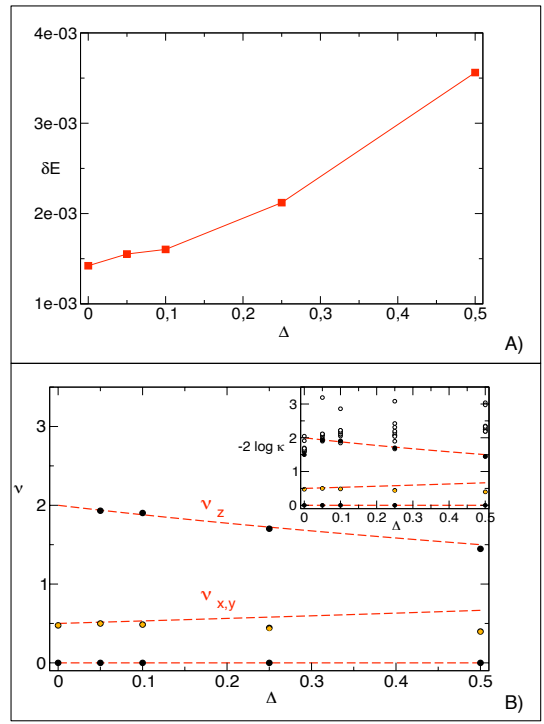}
\caption{Energy error (Part A) and critical exponents (Part B) of the XXZ model for 
various values of $\Delta$. For comparison in the inset we report also the other eigenvalues of the map.
(plots obtained for $m=4$).}
\label{XXY-bmark}
\end{center}
\end{figure}

\begin{figure}[t!]
\begin{center}
\includegraphics[scale=0.4]{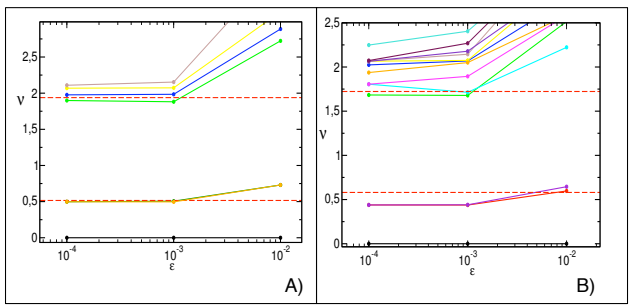}
\caption{Convergence of the cirtical exponents for the $XXY$ model for
$\Delta = 0.05$ (left) and $\Delta = 0.1$ (right). Dashed lines gives the
exact critical exponents $\nu_\alpha$.}
\label{XXY-bmark-1}
\end{center}
\end{figure}

 The results obtained for the Ising model are
very promising, however it is well known that the Ising model has a
simple spectrum and properties due to the fact that it is equivalent
to free fermions~\cite{lieb61}. We then afford the study of a more
complex model, the $XXZ$ model. In Fig.~\ref{XXY-bmark} (Part A) we show the
error of the energy ground state at the thermodynamic limit for
various values of $\Delta$ and $m=4$. As expected, the error is
increasing with $\Delta$, however it remains of the order of 
$\delta E = 10^{-3}$. In Fig.~\ref{XXY-bmark} (Part B) we show the correspondent
critical exponents. As it can be seen, the unital eigenvalue is always
present (at almost machine precision) while the first two degenerate
eigenvalues are related to the $\nu_x, \nu_y$ critical exponents 
(black circle and orange star) with errors increasing with $\Delta$ 
reflecting the precision in the energy of the ground state.
The third critical exponent is also detected with even 
a better precision than the first two for $\Delta > 0.1$ . 
In the case $\Delta=0$ the eigenvalue correspondent to $\nu_z$ does
not happen to be the third one as a few more are present between the
``physical'' ones. This behavior should be investigated in more
details (the prefactors could be zero). 
Finally we report in Fig.~\ref{XXY-bmark-1} the convergence of the
critical exponents for the $XXZ$ model for different values  of $\Delta$ as a function of the 
maximum ``move'' in the minimization procedure $\epsilon$ which corresponds
to the convergence of the ground state.
 As can clearly seen, the behavior is regular, converging to a
value that is near to the expected one.

\section{Conclusions} \label{conclusions}

This paper presents a recursive algorithm for optimizing the MERA tensor description of the ground state $|\Psi_G\rangle$ of a critical many-body Hamiltonian in the thermodynamic
limit of infinitely many sites. It is based on the results of Ref.~\cite{quMERA} which established a connection between $|\Psi_G\rangle$ and  the transfer super-operator of the
 MERA network that approximates it.

 While completing the writing of the present manuscript 
 we became aware of a work  by Pfeifer, Evenbly and Vidal 
(arXiv:0810.0580v1)
 employing a modified MERA tensor network representation  seems to lead to a more 
 efficient evaluation of the critical exponents.

\acknowledgments 
This work was in part founded by the Quantum Information
research program of Centro di Ricerca Matematica Ennio De Giorgi
of Scuola Normale Superiore.


\begin{thebibliography}{99}
\bibitem{NOTA11} 
 As discussed in Ref.~\cite{quMERA}, generic homogeneous MERA are characterized
 by  two QuMERA channels $\Phi_R$ and $\Phi_L$. However if the MERA has to describe a translational
  invariant system the two will, by necessity, have the same fix point and the same spectrum. Without loss 
  of generality we can thus define $\Phi$ as the convex combination of $\Phi_{R,L}$, i.e. $\Phi := (\Phi_R + \Phi_L)/2$.
\end{thebibliography}
\end{document}